\documentclass[12pt]{article}
\usepackage{times}

\usepackage{epsf}
\def\beq{\begin{equation}}
\def\eeq{\end{equation}}
\def\bea{\begin{eqnarray}}
\def\eea{\end{eqnarray}}
\def\ba{\begin{array}}
\def\ea{\end{array}}
\def\bea{\begin{eqnarray}}
\def\eea{\end{eqnarray}}

\begin{document}
\title{ Exact solutions of Schr\"{o}dinger equation for PT-/non-PT-symmetric
and non-Hermitian Exponential Type Potentials with the
position-dependent effective mass }

\author{\"Ozlem Ye\c{s}ilta\c{s} $^1$ and Ramazan Sever $^2$
         \thanks{Corresponding author: sever@metu.edu.tr}\\
$^1$ Turkish Atomic Energy Authority, Istanbul Road, 30 km Kazan 06983, Ankara, Turkey \\
$^2$ Department of Physics, Middle East Technical University, 06531,
Ankara, Turkey}
\date{\today}
\maketitle
\normalsize

\begin{abstract}
\noindent Exact solutions of Schr\"{o}dinger equation for
PT-/non-PT-symmetric and non-Hermitian Morse  and P\"{o}schl-Teller
potentials are obtained with the position-dependent effective mass
by applying a point canonical transformation method. Three kinds of
mass distributions are used in order to construct exactly solvable
target potentials and obtain energy spectrum and corresponding wave
functions.
\end{abstract}

\newpage

\section{Introduction}

\noindent In the past few years, theoretical researches on great
variety of non-Hermitian Hamiltonians have received an important
increase. Because many of these systems are invariant under combined
parity and time reversal (PT) transformation which lead to either
real (in case of broken PT symmetry) or pairs of complex conjugate
energy eigenvalues (in case of spontaneously broken PT symmetry)
[1,2]. This property of energy eigenvalues in non-Hermitian PT
invariant systems can be related to the pseudo-hermiticity [3] or
anti-unitary symmetry [4,5] of the corresponding Hamiltonians. In
ref. [6] it was proposed a new class of non-Hermitian Hamiltonians
with real spectra which are obtained using pseudo-symmetry.
Moreover, completeness and orthonormality conditions for eigenstates
of such potentials are proposed [7]. In the study of PT-invariant
potentials various techniques have been applied to a great variety
of quantum mechanical fields as variational methods, numerical
approaches, Fourier analysis, semi-classical estimates, quantum
field theory and Lie group theoretical approaches [7-16]. In
additional, PT-symmetric and non-PT symmetric and also non-Hermitian
potential cases such as oscillator type potentials and a variety of
potentials within the framework of SUSYQM [17-21], exponential type
screened potentials [22], quasi/conditionally exactly solvable ones
[23], PT-symmetric and non-PT symmetric and also non-Hermitian
potential cases within the framework of SUSYQM via Hamiltonian
Hierarchy Method [24] and some others are studied [25-27].
\\

\noindent On the other hand, there has been respected interest in a
position-dependent mass, which is generally written as (PDM)
$M(r)=m_{0}m(r)$, problems associated with a quantum mechanical
particle forms an effective model for the study of many physical
problems [28-39] due to considerable applications in condensed
matter physics and material science. The model applied to wide
variety of physical systems such as quantum dots [40], liquid
crystals [41], kinetics of evolution of microstructures and atomic
displacements in the string [42], He cluster [43] semiconductor
heterostructures [44] and nuclei [45]. Generally, those works are
concentrated in obtaining the energy eigenvalues and the potential
function for the given quantum system with the PDM. In the mapping
of nonconstant mass Schr\"{o}dinger equation, point canonical
transformations (PCTs) are employed [46-49]. During the process, it
is needed to transform non-constant mass, which is known as
"effective mass" characterizes the curvature of the dispersion
relation, to a constant one so that the latter equation can be
solved. Hence, energy spectra and corresponding wave functions of
the target problem are produced easily. Various potentials, which
satisfy the concept of exactly solvability, such as oscillator,
Coulomb, Morse [50], hard-core potential [51], trigonometric type
[52] and conditionally exactly solvable potentials [53] as well as
the Scarf and Rosen-Morse type [54] ones including the PT-symmetry
are considered for the construction of exact solution via PCT. The
aim of this work is to apply PCT to the exact solutions of the
nonconstant mass Schr\"{o}dinger equation for P\"{o}schl-Teller and
Morse potentials which are complex and/or PT/non-PT symmetric,
non-Hermitian and the exponential type systems.
\\
\\
\noindent The contents of the present paper is as follows: In
section II, it is shown how to construct effective mass
Schr\"{o}dinger equation by using PCT method. In section III, IV and
V, using three different type mass distributions, PCT method is
applied to general Morse and non-Hermitian, PT/Non-PT symmetric
Morse potentials. In section VI, VII and VIII, the general form of
P\"{o}schl-Teller potential and non-Hermitian, PT/Non-PT symmetric
P\"{o}schl-Teller potentials are studied by using PCT method  within
three different mass functions in order to construct the target
problem including energy eigenvalues and corresponding wavefunctions
within PT symmetry.
\\
\section{Effective Mass Schr\"{o}dinger equation }

\noindent As is well known, the general form of one dimensional time
independent position-dependent mass Schr\"{o}dinger equation (PDMSE)
gives rise to

\begin{eqnarray}
-\frac{1}{2}\left[\nabla_{x}\frac{1}{M(x)}\nabla_{x}\right]\psi(x)-
\left[E-V(x)\right]\psi(x)=0,
\end{eqnarray}

\noindent where $M(x)=m_{0}m(x)$. So the Eq. (1) reads

\begin{eqnarray}
\psi^{''}(x)-\left(\frac{m^{'}}{m}\right)\psi^{'}(x)+2m\left[E-V(x)\right]\psi(x)=0,
\end{eqnarray}

\noindent where $\hbar=1$ amd $m_{0}$ is a constant. The one
dimensional Schr\"{o}dinger equation with a constant mass is

\begin{eqnarray}
\Phi^{''}(y)+2\left[\varepsilon-V(y)\right]\Phi(y)=0.
\end{eqnarray}

\noindent A transformation is defined as $y\rightarrow x$ and for a
mapping $y=f(x)$, we rewrite the wave functions in the form

\begin{eqnarray}
\Phi(y)=g(x)\psi(x)
\end{eqnarray}

\noindent The transformed Schr\"{o}dinger equation reads

\begin{eqnarray}
\psi^{''}(x)+2\left(\frac{g^{'}}{g}-\frac{f^{''}}{f^{'}}\frac{g^{'}}{g}\right)\psi^{'}(x)+
\left(\left(\frac{g^{''}}{g}-\frac{f^{''}}{f^{'}}\frac{g^{'}}{g}\right)+
2(f^{'})^{2}\left[V(f(x)-\varepsilon)\right]\right) \psi(x)=0.
\end{eqnarray}

\noindent Comparing Eqs. (2) and  (5), we get the following
identities

\begin{eqnarray}
g(x)=\sqrt{\frac{f^{'}(x)}{m(x)}}
\end{eqnarray}

\noindent and

\begin{eqnarray}
V(x)-E=\frac{(f^{'})^{2}}{m}\left[V(f(x)-\varepsilon)\right]-\frac{1}{2m}F(f,g)
\end{eqnarray}

\noindent where $F(f,g)=\left(\frac{g^{''}}{g}-\frac{f^{''}}{f^{'}}
\frac{g^{'}}{g}\right)$. As it is seen from  Eqs. (2) and (5), if we
substitute $(f^{'})^{2}=m$ in Eq. (7), then the reference problem is
transformed to the target problem including the energy spectra of
the bound states, potential and wave function as

\begin{eqnarray}
E_{n}=\varepsilon_{n}
\end{eqnarray}

\begin{eqnarray}
V(x)=V(f(x))-\frac{1}{8m}\left[\frac{m^{''}}{m}-
\frac{7}{4}\left(\frac{m^{'}}{m}\right)^{2}\right]
\end{eqnarray}

\begin{eqnarray}
\psi(x)= [m(x)]^{1/4} \Phi_{n}(f(x)).
\end{eqnarray}

\noindent The PCT method can be applied to a problem which has an
exact solution by using the procedure given below.

\section{Generalized Morse Potential}

\noindent Consider the Morse potential as the reference problem
[19,22]

\begin{eqnarray}
V(y)=V_{1}e^{-2\alpha y}-V_{2}e^{-\alpha y}
\end{eqnarray}

\noindent The energy eigenvalues and eigenfunctions of the our
source potential are given as

\begin{eqnarray}
\varepsilon_{n}=-\frac{\alpha^{2}}{4}\left[\frac{V_{2}}{\alpha
\sqrt{V_{1}}}-(2n+1)\right]^{2}
\end{eqnarray}

\begin{eqnarray}
\Phi_{n}(y)=C_{n}s^{2\epsilon}e^{-\gamma s}L^{4\epsilon}_{n}(2\gamma
s)
\end{eqnarray}

\noindent where $s=\sqrt{V_{1}}e^{-\alpha y}$.

\subsection{Asymptotically vanishing mass distribution}

\noindent In this section, we use asymptotically vanishing type mass
distribution as given below in order to get some target potentials
providing us the exact solutions

\begin{eqnarray}
m(x)=\frac{\alpha^{2}}{x^{2}+q}
\end{eqnarray}

\noindent The mapping function becomes

\begin{eqnarray}
y=f(x)=\int^{x} \sqrt{m(x)}dx=\alpha ln
\left(x+\sqrt{x^{2}+q}\right)
\end{eqnarray}

\noindent and

\begin{eqnarray}
x=sinh_{q}\left(\frac{y}{\alpha}\right), \alpha\neq0.
\end{eqnarray}

\noindent Using Eqs. (12-16), the new potential is obtained as

\begin{eqnarray}
V(x)=V_{1}\left(x+\sqrt{x^{2}+q}\right)^{-2\alpha^{2}}-
V_{2}\left(x+\sqrt{x^{2}+q}\right)^{-\alpha^{2}}-
\frac{1}{8\alpha^{2}}\left(1+\frac{q}{x^{2}+q}\right)
\end{eqnarray}

\noindent Hence, the energy eigenvalues and corresponding wave
functions for the general Morse potential are obtained as

\begin{eqnarray}
E_{n}=\varepsilon_{n}
\end{eqnarray}

\noindent and

\begin{eqnarray}
\psi_{n}(x)=C_{n}\frac{\sqrt{\alpha}}{\left(x^{2}+
q\right)^{1/4}}(f(x))^{2\epsilon} e^{-\gamma f(x)}
L^{4\epsilon}_{n}(2\gamma f(x)).
\end{eqnarray}

\noindent where $\epsilon^{2}=-\frac{E}{2\alpha^{2}}$,
$\gamma=\frac{1}{\alpha^{2}}$.

\subsection{Mass Distribution $m(x)=\frac{\alpha^{2}}{(b+x^{2})^{2}}$}

\noindent In the second example of mass distribution, the mapping
function becomes

\begin{eqnarray}
y=f(x)=\alpha tan^{-1}\frac{x}{q}
\end{eqnarray}

\noindent and

\begin{eqnarray}
x=\sqrt{q}tan(\frac{y}{\alpha})
\end{eqnarray}

\noindent The mapping function leads to the following target system
having the same energy spectra

\begin{eqnarray}
V(x)=V_{1}e^{-2\alpha^{2}tan^{-1}\frac{x}{\sqrt{q}}}-
V_{2}e^{-\alpha^{2}tan^{-1}\frac{x}{\sqrt{q}}}-
\frac{73x^{2}-16q}{32\alpha^{2}}
\end{eqnarray}

\noindent and

\begin{eqnarray}
\psi_{n}(x)=\sqrt{\frac{q+x^{2}}{\alpha}}(f(x))^{2\epsilon}
e^{-\gamma f(x)} L^{4\epsilon}_{n}(2\gamma f(x)).
\end{eqnarray}

\subsection{Exponential Type Mass Distribution}

\noindent If we consider the third type exponential mass function
given as

\begin{eqnarray}
m(x)=e^{-\alpha x}
\end{eqnarray}

\begin{eqnarray}
y=f(x)=-\frac{2}{\alpha}e^{-\frac{\alpha x}{2}}
\end{eqnarray}

\noindent and

\begin{eqnarray}
x=-\frac{2}{\alpha}ln(-\frac{\alpha y}{2})
\end{eqnarray}

\noindent The mapping yields to potential with the same energy
spectra

\begin{eqnarray}
V(x)=V_{1}e^{4e^{-\alpha x/2}}-V_{2}e^{2e^{-\alpha
x/2}}+\frac{3\alpha^{2}e^{-\alpha x}}{32}.
\end{eqnarray}

\noindent and corresponding wave function is

\begin{eqnarray}
\psi_{n}(x)=(f(x))^{2\epsilon}e^{-\alpha x/4-\gamma
f(x)}L^{4\epsilon}_{n}(2\gamma f(x))
\end{eqnarray}

\section{Non-PT symmetric and non-Hermitian Morse Potential}

\noindent In the equation (11), if the potential parameters are
defined as $V_{1}=(A+iB)^{2}$, $V_{2}=(2C+1)(A+iB)$ and $\alpha=1$,
then the potential becomes

\begin{eqnarray}
V(y)=(A+iB)^{2}e^{-2y}-(2C+1)(A+iB)e^{-y}
\end{eqnarray}

\noindent where $A$, $B$ and $C$ are arbitrary real parameters and
$i=\sqrt{-1}$. Similarly, the energy eigenvalues for the reference
potential is given as [19,22]

\begin{eqnarray}
\varepsilon_{n}=-(n-C)^{2}
\end{eqnarray}

\subsection{Asymptotically vanishing mass distribution}

\noindent Following the same procedure as in above, we get the
target system

\begin{eqnarray}
E_{n}=\varepsilon_{n}
\end{eqnarray}

\begin{eqnarray}
V(x)=(A+iB)^{2}\left(x+\sqrt{x^{2}+q}\right)^{-2 \alpha}-
(2C+1)(A+iB)\left(x+\sqrt{x^{2}+q}\right)^{- \alpha}-
\frac{1}{8\alpha^{2}}\left(1+\frac{q}{x^{2}+q}\right).
\end{eqnarray}

\subsection{Mass Distribution $m(x)=\frac{\alpha^{2}}{(b+x^{2})^{2}}$}

\noindent Following the same procedure, we obtain the target
potential with same energy spectra and

\begin{eqnarray}
V(x)=(A+iB)^{2}e^{-2\alpha tan^{-1}\frac{x}{\sqrt{q}}}-
(2C+1)(A+iB)e^{-\alpha tan^{-1}\frac{x}{\sqrt{q}}}-
\frac{73x^{2}-16q}{32\alpha^{2}}
\end{eqnarray}

\subsection{Exponential Type Mass Distribution}

\noindent We obtain the target potential with same energy spectrum
for exponential type mapping function as

\begin{eqnarray}
V(x)=(A+iB)^{2}e^{\frac{4e^{-\frac{\alpha x}{2}}}{\alpha}}-
(2C+1)(A+iB)e^{\frac{2e^{-\frac{\alpha
x}{2}}}{\alpha}}+\frac{3\alpha^{2}e^{-\alpha x}}{32}.
\end{eqnarray}

\section{PT symmetric and non-Hermitian Morse Potential}

\noindent When $\alpha=i\alpha$ and $V_{1}, V_{2}$ are real, the Morse potential
becomes

\begin{eqnarray}
V(y)=V_{1}e^{-2i \alpha y}-V_{2}e^{-i\alpha y}
\end{eqnarray}

\noindent The energy eigenvalues are given for this potential as
[19,22]

\begin{eqnarray}
\varepsilon_{n}=\alpha^{4}\left[(n+\frac{1}{2})+\frac{V_{2}}{2\alpha
\sqrt{|-V_{1}|}}\right]^{2}
\end{eqnarray}

\noindent If we take the parameters of Eq.(25) as
$V_{1}=-\omega^{2}$, $V_{2}=D$ and $\alpha=2$ then, corresponding
energy eigenvalues for any n-th state are,

\begin{eqnarray}
\varepsilon_{n}=(2n+1+\frac{D}{2\omega})^{2}
\end{eqnarray}

\noindent which ,is studied by Znojil and Bagchi and Quesne.
[10-11,19].

\subsection{Asymptotically vanishing mass distribution}

\noindent  Thus, the target system with asymptotically vanishing
mass distribution are given as

\begin{eqnarray}
E_{n}=\varepsilon_{n}
\end{eqnarray}

\begin{eqnarray}
V(x)=V_{1}\left(x+\sqrt{x^{2}+q}\right)^{-2i \alpha^{2}}-
V_{2}\left(x+\sqrt{x^{2}+q}\right)^{-i \alpha^{2}}-
\frac{1}{8\alpha^{2}}\left(1+\frac{q}{x^{2}+q}\right).
\end{eqnarray}

\subsection{Mass Distribution $m(x)=\frac{\alpha^{2}}{(b+x^{2})^{2}}$}
\noindent In the PT symmetric and non-Hermitian case, new potential
is given by

\begin{eqnarray}
V(x)=V_{1}\sqrt{\frac{1-2\alpha^{2}\frac{x}{\sqrt{q}}}{1+2\alpha^{2}\frac{x}{\sqrt{q}}}}-
V_{2}\sqrt{\frac{1-\alpha^{2}\frac{x}{\sqrt{q}}}{1+\alpha^{2}\frac{x}{\sqrt{q}}}}-
\frac{73x^{2}-16q}{32\alpha^{2}}.
\end{eqnarray}

\subsection{Exponential Type Mass Distribution}
\noindent The new potential with the same energy spectra is

\begin{eqnarray}
V(x)=V_{1}e^{4ie^{-\alpha x/2}}-V_{2}e^{2ie^{-\alpha
x/2}}+\frac{3\alpha^{2}e^{-\alpha x}}{32}.
\end{eqnarray}

\section{P\"{o}schl-Teller Potential}

\noindent The general form of the P\"{o}schl-Teller potential is
[19,22]

\begin{eqnarray}
V(y)=-4V_{0}\frac{e^{-2\alpha y}}{(1+qe^{-2\alpha y})^{2}}
\end{eqnarray}

\noindent Its energy spectra and corresponding wavefunctions are

\begin{eqnarray}
\varepsilon_{n}=-\frac{\alpha^{2}}{4}
\left(-(2n+1)+\sqrt{1+\frac{8V_{0}}{q\alpha^{2}}}\right)^{2}
\end{eqnarray}

\begin{eqnarray}
\psi_{n}(y)=s^{-\epsilon}(1-s)^{\nu/2}P^{(2\epsilon,\nu-1)}_{n}(1-2qs).
\end{eqnarray}

\noindent where $s=-e^{-2\alpha y }$,
$P^{-\nu_{2}-\frac{1}{2},\nu_{2}-\frac{1}{2}}_{n}(y)$ stands for
Jacobi polynomials and
$\nu_{1}=\sqrt{1+\frac{8V_{0}}{q\alpha^{2}}}$,\,\,
$\nu_{2}=\sqrt{\frac{8V_{0}}{q\alpha^{2}}}$.

\subsection{Asymptotically vanishing mass distribution}

\noindent The target system is obtained as with the mass function

\begin{eqnarray}
E_{n}=\varepsilon_{n}
\end{eqnarray}

\begin{eqnarray}
V(x)=-4V_{0}\frac{\left(x+\sqrt{x^{2}+q}\right)^{-2\alpha^{2}}}{\left[1+q(\left(x+
\sqrt{x^{2}+q}\right)^{-2\alpha^{2}}\right]^{2}}-
\frac{1}{8\alpha^{2}}\left(1+\frac{q}{x^{2}+q}\right)
\end{eqnarray}

\begin{eqnarray}
\psi_{n}(x)=\frac{\left(x^{2}+
q\right)^{1/4}}{\sqrt{\alpha}}(f(x))^{-\epsilon}(1-f(x))^{\nu/2}P^{(2\epsilon,\nu-1)}_{n}(1-2qf(x))
\end{eqnarray}

\subsection{Mass Distribution $m(x)=\frac{\alpha^{2}}{(q+x^{2})^{2}}$}
\noindent If we consider to obtain a target potential for the
P\"{o}schl-Teller Potential, it can be obtain as

\begin{eqnarray}
V(x)=-4V_{0}\frac{e^{-2\alpha^{2}tan^{-1}\frac{x}{\sqrt{q}}}}{\left(1+
qe^{-2\alpha^{2}tan^{-1}\frac{x}{\sqrt{q}}}\right)^{2}}-
\frac{73x^{2}-16q}{32\alpha^{2}}.
\end{eqnarray}

\begin{eqnarray}
\psi_{n}(x)=\frac{\left(x^{2}+
q\right)^{1/2}}{\sqrt{\alpha}}(f(x))^{-\epsilon}(1-f(x))^{\nu/2}P^{(2\epsilon,\nu-1)}_{n}(1-2qf(x))
\end{eqnarray}
\subsection{Exponential Type Mass Distribution}
\noindent With the exponential Type Mass Distribution, it can be
obtained with the same energy spectra as

\begin{eqnarray}
V(x)=-4V_{0}\frac{e^{4e^{-\alpha x/2}}}{[1+qe^{4e^{-\alpha
x/2}}]^{2}}+\frac{3\alpha^{2}e^{-\alpha x}}{32}.
\end{eqnarray}

\noindent and

\begin{eqnarray}
\psi_{n}(x)=e^{\alpha
x/4}(f(x))^{-\epsilon}(1-f(x))^{\nu/2}P^{(2\epsilon,\nu-1)}_{n}(1-2qf(x))
\end{eqnarray}

\section{Non-PT symmetric and non-Hermitian P\"{o}schl-Teller
cases}

\noindent In this case, $V_{0}$ and $q$ are complex parameters
$V_{0}=V_{0R}+iV_{0I}$ and $q=q_{R}+iq_{I}$ but $\alpha$ is a real
parameter. Although the potential is complex and the corresponding
Hamiltonian is non-Hermitian and also non-PT symmetric, there may be
real spectra if and only if $V_{0I}q_{R}=V_{0R}q_{I}$. When both
parameters $V_{0}$ and $q$ are taken pure imaginary, the potential
turns out to be [19,22],

\begin{eqnarray}
V(y)=-4V_{0}\frac{2qe^{-4\alpha y}+i(1-q^{2}e^{-4\alpha
y})}{(1+q^{2}e^{-4\alpha y})^{2}}
\end{eqnarray}

\noindent For simplicity, we use the notation $V_{0}$ and $q$
instead of $V_{0I}$ and $q_{I}$. In this case,  the same energy
eigenvalues are obtained as in the Eq.(30).

\subsection{Asymptotically vanishing mass distribution}

\noindent  The new potential is

\begin{eqnarray}
V(x)=-4V_{0}\frac{2q\left(x+\sqrt{x^{2}+q}\right)^{-4\alpha^{2}}+
i\left(1-q^{2}\left(x+\sqrt{x^{2}+q}\right)^{-4\alpha^{2}}\right)}{\left[1+q^{2}\left(x+\sqrt{x^{2}+q}\right)^
{-4\alpha^{2}}\right]^{2}}-
\frac{1}{8\alpha^{2}}\left(1+\frac{q}{x^{2}+q}\right).
\end{eqnarray}

\subsection{Mass Distribution
$m(x)=\frac{\alpha^{2}}{(b+x^{2})^{2}}$} \noindent The target
potential for this case can be obtained as

\begin{eqnarray}
V(x)=-4V_{0} \frac{2qe^{-4\alpha^{2}tan^{-1}\frac{x}{\sqrt{q}}}+
i\left(1-q^{2}e^{-4\alpha^{2}tan^{-1}\frac{x}{\sqrt{q}}}\right)}{\left(1+q^{2}
e^{-4\alpha^{2}tan^{-1}\frac{x}{\sqrt{q}}}\right)^{2}}-
\frac{73x^{2}-16q}{32\alpha^{2}}.
\end{eqnarray}
\subsection{Exponential Type Mass Distribution}
\noindent The target potential for this case can be obtained with
same the energy eigenvalues as

\begin{eqnarray}
V(x)=-4V_{0} \frac{2qe^{8e^{-\alpha x/2}}+i(1-q^{2})e^{8e^{-\alpha
x/2}}}{(1+q^{2}e^{8e^{-\alpha
x/2}})^{2}}+\frac{3\alpha^{2}e^{-\alpha x}}{32}.
\end{eqnarray}

\section{PT symmetric and non-Hermitian  P\"{o}schl-Teller cases}

\noindent We choose the parameters $V_{0}$ and$q$ are real and also
$\alpha=i \alpha$. Then, the potential turns into

\begin{eqnarray}
V(x)=-4V_{0}\frac{(1+q^{2})cos2\alpha x+2q+i(q^{2}-1)sin2\alpha
x}{(1+q^{2})^{2}+4q cos2\alpha x((1+q cos2\alpha x+q^{2})}
\end{eqnarray}

\noindent and corresponding energy eigenvalue is given as [19,22]

\begin{eqnarray}
\varepsilon_{n}=-\frac{\alpha^{2}}{4}\left[2n+1+\sqrt{1+\frac{16V_{0}}{\alpha^{2}}}\right]^{2}
\end{eqnarray}

\subsection{Asymptotically vanishing mass distribution}

\noindent  The new system is

\begin{eqnarray}
E_{n}=\varepsilon_{n}
\end{eqnarray}

\begin{eqnarray}
V(x)
&=-4V_{0}\frac{\left[q\left(x+\sqrt{q+x^{2}}\right)^{i\alpha^{2}}-
\left(x+\sqrt{q+x^{2}}\right)^{-i\alpha^{2}}\right]^{2}}{\left(1+q^{2}\right)^{2}+
4qcos\left[2\alpha^{2}ln\left(x+\sqrt{q+x^{2}}\right)\right]
\left(1+qcos\left[2\alpha^{2}ln\left(x+\sqrt{q+x^{2}}\right)\right]+q^{2}\right)}-
\frac{1}{8\alpha^{2}}\left(1+\frac{q}{x^{2}+q}\right)
\end{eqnarray}

\subsection{Mass Distribution $m(x)=\frac{\alpha^{2}}{(b+x^{2})^{2}}$}

\noindent The new potential is given as

\begin{eqnarray}
V(x) =-4V_{0}\frac{(qe^{i\alpha^{2}tan^{-1}\frac{x}{\sqrt{q}}}+
e^{-i\alpha^{2}tan^{-1}\frac{x}{\sqrt{q}}})^{2}}{(1+q^{2})^{2}+4qcos(2
\alpha^{2}tan^{-1}\frac{x}{\sqrt{q}}))(1+qcos2\alpha^{2}tan^{-1}\frac{x}{\sqrt{q}}+q^{2})}-
\frac{73x^{2}-16q}{32\alpha^{2}}.
\end{eqnarray}
\subsection{Exponential Type Mass Distribution}
\noindent The target potential can be obtained with having the same
energy spectra for this case as

\begin{eqnarray}
V(x) =-4V_{0}\frac{(qe^{-2ie^{-\alpha x/2}}+e^{2ie^{-\alpha
x/2}})^{2}}{(1+q^{2})^{2}+4qcos(4e^{-\alpha x/2})(1+q
cos(4e^{-\alpha x/2})+q^{2})}+\frac{3\alpha^{2}e^{-\alpha x}}{32}.
\end{eqnarray}

\section{Conclusions}

\noindent In this article we have explored the PCT approach to a
class of exponential type PT/non-PT symmetric and nonhermitian
Hamiltonians such as Morse and P\"{o}schl-Teller potentials with
some spatially dependent effective masses. For each type of
potentials we have obtained a set of exactly solvable target
potentials by using three position dependent mass distributions. It
is pointed out that the importance of the mapping function which
aids to construct of closed forms of the energy spectrum and
corresponding wavefunctions. Specially, the reference system with
the source potential and the new system with the target potential
have the same  bound state energy eigenvalues. It was shown that the
exact solvability depends not only on the form of the potential, but
also on the spatial dependence on the mass within the PT symmetric
framework.

\section{Acknowledgements}

This research was partially supported by the Scientific and
Technological Research Council of Turkey.

\end{document}